\documentclass[12pt]{article}
\usepackage{graphicx}
\begin{document}
\begin{center}
{\large\bf 3- and 4- body meson- nuclear clusters} \\[4mm]
{V. B. Belyaev$^a$, W. Sandhas$^b$, and I. I. Shlyk$^a$}\\
{\small \em $^a$ Joint Institute for Nuclear Research, BLTP,
Dubna, 141980, Russia, \\ $^b$ Physikalisches Institut,
Universitat Bonn, D-53115 Bonn, Germany}
\end{center}

\begin{abstract}
The binding energies and matter distributions for the 3- body
systems like $\phi$- meson $ + 2N$, $2\phi + N$ and 4- body system
like $\phi+3n$ are calculated. For the 3- particle systems two-
dimensional Faddeev equations in the differential form are used.
For the 4- body system $\phi+3n$ the folding model is
applied.
\end{abstract}
\section{Introduction}
    As it has recently been intensively discussed [1-7], there are
    indications of strong attraction of mesons with one
    strange quark $K^-\,$($\,\overline{K}\,$) to few- nucleon
    nuclei. Along this line, it is interesting to look at the
    interaction of a meson with two strange quarks like the
    $\phi$-meson with light nuclei.
    Already existing theoretical investigations of the $\phi$-meson
interaction show rather strong attraction between a $\phi$-meson
and a nucleon. Indeed, the calculation of the $\phi-N$ interaction
within the quark model~\cite{Huang} and on the basis of a totally
different phenomenological model~\cite{Gao} based on the dominant
role of the $s\overline{s}$ configuration in the $\phi$-meson
structure predicts considerable $\phi-N$ attraction with a
binding energy of about 9 MeV for the $\phi N$ system.

Strong attraction is not very surprising in view of the
physical arguments that the strong $K^-N$ attraction originates 
from the influence of subthreshold resonances
$\Lambda_{\,1405}$ and $\Sigma_{\,1385}$.

Indeed, let us compare the mass of the state $\phi + N$ with the
masses of two subthreshold states $K + \Lambda_{\,1405}$ and $K +
\Sigma_{\,1385}$. It turned out that distances of the above
subthreshold states from the $\phi + N$ threshold are of the same
order of magnitude as in the $K^-N$ case, which means that as in
$K^-N$ one can expect strong influence of $\Lambda_{\,1405}$ and
$\Sigma_{\,1385}$ and strong attraction also in the $\phi N$
system.

Bearing in mind this sort of strong attraction in the $\phi N$
system, it is interesting to consider the possibility of bound
states of a $\phi$-meson with a few nucleons, in particular, with two
neutrons or two protons. This is in fact a question concerning the
 existence of new nuclear clusters which do not exist without a
 $\phi$- meson. In what follows, we will calculate the binding energies 
 of the three-body systems $\phi n n$, $\phi n p$ and $\phi p p$.

\section{Input}
It is reasonable to consider a 3- particle system of the type 
$\phi \phi N$ in the same theoretical frameworks. To describe this
system, one needs apart from the $\phi-N$ potential also the
$\phi-\phi$ potential. One of the simplest ways to construct
this potential is to take it in the form of a sum of two Yukawa
potentials (attractive and repulsive):
\begin{equation}
\label{phi-phi potential} V_{\,\phi
\phi}(r)\;=\;V_{\,1}\,\frac{\,e^{-\mu_{\,1} r} }
{\,r}\,-\,V_{\,2}\,\frac{\,e^{-\mu_{\,2}\, r} } {\,r}
\end{equation}
where
$$ V_{\,1}\;=\; 1000\;\mathrm{MeV} \cdot
\mathrm{fm}\;,\;\;\mu_{\,1}\;=\;2.5 \;\mathrm{fm}^{\,-1}\;,
$$
$$ V_{\,2}\;=\; 1250\;\mathrm{MeV} \cdot
\mathrm{fm}\;,\;\;\mu_{\,2}\;=\;3.0 \;\mathrm{fm}^{\,-1}\;.
$$

\begin{figure}
\begin{center}
{\includegraphics[angle=0,width=11cm]{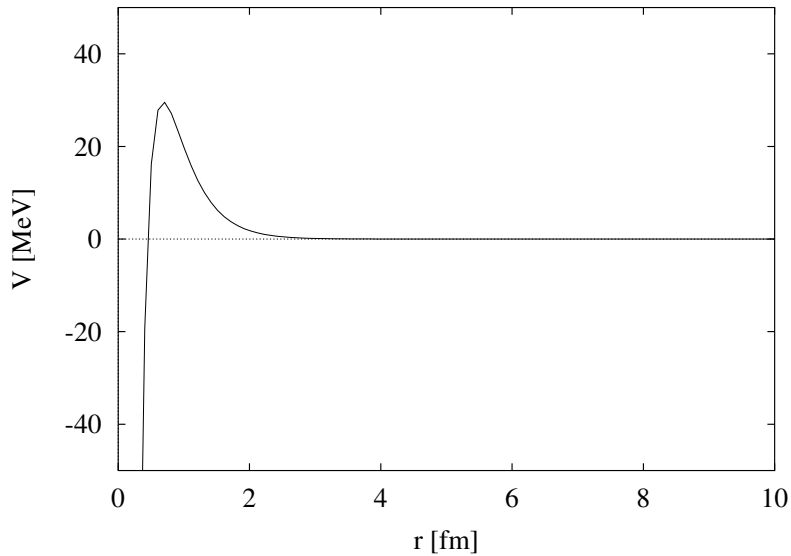}} \quad\mbox{}
\end{center}
\vspace{-7mm} \caption{The $\phi \phi$ local potential} \label{phi-phi
potential}
\end{figure}

The parameters of this potential were fixed by the position and
width of the f2(2010) resonance which has only one decay channel
into two $\phi$- mesons.

\begin{figure}
\begin{center}
{\includegraphics[angle=0,width=11cm]{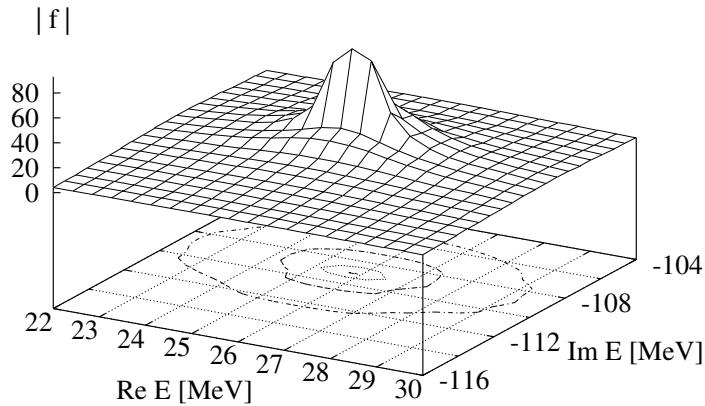}} \quad\mbox{}
\end{center}
\vspace{-7mm} \caption{The location of the pole of the absolute value
of the $\phi - \phi$ scattering amplitude in the complex energy
plain. } \label{resonance}
\end{figure}

As in~\cite{Gao}, a Yukawa type potential is chosen for the
$\phi-N$ interaction
\begin{equation}
\label{Yukawa phi N} V_{\,\phi N}(r)\;=\;-\,\alpha\,e^{\,-\,\mu\,
r}/r
\end{equation}
with $\alpha=1.25$ and $\mu=600$ MeV. This potential is rather
deep and narrow and supports binding in the $\phi N$ system with
binding energies $E_{\,\phi n}=-9.47$ and $E_{\,\phi p}=-9.40$
MeV.

For the $np$ triplet s-wave interaction the potential MT
III~\cite{Tjon} was used. Our singlet s-wave interaction is
based on the potential MT I~\cite{Tjon} with a slight
modification of having now a parameter $\lambda_{\,A}=2.617$. This
value is chosen in order to
 reproduce the
experimental value of the $nn$-scattering length $a_{\,nn}=-18.5$
fm~\cite{Howell}. One can see that in the three-body systems $\phi
N N$ there are two scales of distance related to the different
ranges of the $N-N$ and $\phi-N$ interactions. This may produce a
delicate interplay between a narrow attraction area of the
$\phi-N$ interaction and repulsive parts of the MT-potentials, as
it was emphasized in~\cite{our in ArXiv}. Apart from that,
different ranges of the interaction can provide the cluster
formation in the systems under consideration.

Let us start to describe three particle systems.

\section{Calculations}
Our calculations are based on Faddeev equations~\cite{Faddeev} in a 
differential form~\cite{Pupyshev} written down for the 3-body
systems $\phi N N$ and $\phi \phi N$.

First, the Faddeev components of the wave function are expanded
into partial waves:
\begin{equation}
\Psi_{\,\alpha}(\vec{\eta}_{\,\alpha},
\vec{\xi}_{\,\alpha})\;=\;\sum_{\,LM\, l\, \lambda}^{\,}\,
\frac{\,1}{\,\eta_{\,\alpha}\,\xi_{\,\alpha}}\,U_{\,\alpha \, l
 \lambda}^{\,L} (\eta_{\,\alpha},\,\xi_{\,\alpha}) \, Y_{\,l
 \lambda}^{\,LM}(\hat{\eta}_{\alpha},\hat{\xi}_{\alpha})\;
 \stackrel{L\,=\,l\,=\,\lambda\,=\,0}{\Longrightarrow}\;
 $$
 $$
\frac{\,1}{\,\eta_{\,\alpha}\,\xi_{\,\alpha}}\,U_{\,\alpha }
(\eta_{\,\alpha},\,\xi_{\,\alpha}) \,
Y_{\,00}^{\,00}(\hat{\eta}_{\alpha},\hat{\xi}_{\alpha})
\end{equation}
where $\eta_{\,\alpha}=|\vec{\eta}_{\,\alpha}|$,
$\xi_{\,\alpha}=|\vec{\xi}_{\,\alpha}|$,
$\hat{\eta}_{\alpha}=\vec{\eta}_{\,\alpha} / \,
|\vec{\eta}_{\,\alpha}|$, $\hat{\xi}_{\alpha}=\vec{\xi}_{\,\alpha}
/ \, |\vec{\xi}_{\,\alpha}|$, $Y_{\,l
 \lambda}^{\,LM}$ are the bispherical harmonics. Jacobi coordinates $\vec{\eta}_{\,\alpha},
\vec{\xi}_{\,\alpha}$ were used, and only the lowest partial
wave is taken into account.

The Jacobi coordinates are as usual:
\begin{equation}
\vec{r}_{\,i}\,-\,\vec{r}_{\,j}\;=\;\frac{\,\vec{\eta}_{\,\alpha}}{\,a_{\,\alpha}} \\
\end{equation}
\begin{equation}
\frac{m_{\,i} \, \vec{r}_{\,i} \,+\, m_{\,j} \, \vec{r}_{\,j}
}{m_{\,i} \,+\, m_{\,j}} \,-\, \vec{r}_{\,k}
\;=\;\frac{\,\vec{\xi}_{\,\alpha} }{\,b_{\,\alpha} }
\end{equation}
 where $\vec{r}_{\,i}$, $ m_{\,i}$ denote the radius-vector and the mass of particle $i$,
$$
a_{\,\alpha}\;=\;\sqrt{\frac{\,m_{\,i} \, m_{\,j} \phantom{t}
}{\,(m_{\,i}\,+\, m_{\,j})\,M}}\;,\;\;
b_{\,\alpha}\;=\;\sqrt{\frac{\,m_{\,k}(m_{\,i} \, + \,
m_{\,j})}{\,M^{\,2}}}\;,\;M\;=\;m_{\,1}\,+\,m_{\,2}\,+\,m_{\,3}\,
$$
and indices $\alpha$ take on the following values: $\,\alpha=3\,$ for
$(i j) k = (12)3$, $\,\alpha=1\,$ for $(i j) k = (23)1$,
$\,\alpha=2\,$ for $(i j) k = (31)2$.

Since there are two identical particles in the system (we take
$m_{\,N}=m_{\,n}$ for $\phi n p$ system), the following two
coupled-differential Faddeev equations survive:
\begin{equation}
\label{Faddeev differential radial} \left\{
\begin{array}{l}
\left[\widehat{D}\;+\;V_{\,1}\left(\frac{\,\rho
\cos{\,\varphi}}{\,a_{\,1}}\right)-\;E\;\right]\,
U_{\,1}(\rho,\,\varphi) \; = \\
\hspace{4cm}-\,V_{\,1}\left(\frac{\,\rho
\cos{\,\varphi}}{\,a_{\,1}}\right) \sum_{\alpha ' \;\neq\;1}
\,\frac{\,1}{\,\sin{(2\, \gamma_{\,\alpha '
1})}}\int_{\,c-}^{\,c+}U_{\,\alpha '}(\rho,\,\varphi ')\,d\varphi
'
\\
\left[\widehat{D}\;+\;V_{\,2}\left(\frac{\,\rho
\cos{\,\varphi}}{\,a_{\,2}}\right)-\;E\;\right]\,
U_{\,2}(\rho,\,\varphi) \; = \\ \hspace{4cm}-\,
V_{\,2}\left(\frac{\,\rho \cos{\,\varphi}}{\,a_{\,2}}\right)
\sum_{\alpha ' \;\neq\;2} \,\frac{\,1}{\,\sin{(2\,
\gamma_{\,\alpha ' 2})}}\int_{\,c-}^{\,c+}U_{\,\alpha
'}(\rho,\,\varphi ')\,d\varphi '
\end{array}
\right.
\end{equation}
$$(U_{\,3} \; \equiv \;U_{\,2})$$ where polar coordinates
 $\rho\;=\;\sqrt{\eta_{\,\alpha}^{\,2}\,+\,\xi_{\,\alpha}^{\,2}}\,
$, $\tan{\varphi_{\,\alpha}}\;=\;\xi_{\,\alpha}/\eta_{\,\alpha}\,$
are introduced and
$$V_{\,1}\;=\;V_{\,NN}\,, \;\;\; \;\;\; V_{\,2}\;=\;V_{\,\phi N}\,,$$
$$\widehat{D}\;=\;-\,\frac{\;\hbar^{\,2}}{\,2M} \left(
\frac{\,\partial^{\,2}}{\,\partial \rho^{\,2}} \; +
\;\frac{\,1}{\,\rho}\,\frac{\,\partial}{\,\partial\rho}\;+
\;\frac{\,1}{\,\rho^{\,2}} \frac{\,\partial^{\,2}}{\,\partial
\varphi^{\,2}} \right)
$$
$$c+\;=\;\mathrm{Min}\left\{|\varphi \,+\, \gamma_{\,\alpha '
\alpha} |\;,\,\pi\,-\,(\varphi \,+\,\gamma_{\,\alpha ' \alpha}
)\right\} \, , \;\;\; \;\;\; c-\;=\;|\varphi\,-\,\gamma_{\,\alpha
' \alpha}|$$
$$\gamma_{\,i j}\;=\;\arcsin{\,s_{\,i j}} \;\, , \;\;\; \;\;\;s_{\,i j}\;=
\;\sqrt{\frac{ \,m_{\,k} \, M } {\,(m_{\,i}\,+\,m_{\,k}) \,
(m_{\,j}\,+\,m_{\,k})} }\,,\; $$
$$(i j k\; =\; 123,\, 231,\, 312)
$$
indices correspond to 1 for $\phi$-meson, 2 and 3 for nucleons.

The two-dimensional system of Faddeev equations~(\ref{Faddeev
differential radial}) has been solved by discretization of
variables: hyperradius $\rho$ and hyperangle $\varphi$ with $N$ and
$M$ mesh points, respectively. Stable results for three digits of
binding energies were reached at $N=110$, $M=210$ and $L(\rho$
variable cutoff$)=9$ fm.

As a result, the binding energy of the system $\phi n n$ with value
$E_{\,\phi n n}=-21.8$ MeV and value $E_{\,\phi
n p}=-37.9$ MeV for the binding of the $\phi n p$ system with $np$
pair in triplet state have been obtained. It should be noticed that for this binding
energy in the $\phi n p$ system both the main $\phi$-meson decay channels
on $K$-mesons are closed. Let us comment the last value of energy
which appeared rather large. From naive reasons in the
configuration $\phi + d$ one would expect binding of an order of $2
\times E_{\phi N}+E_{\,d}$ which is much smaller than the calculated
value. However, due to the strong attraction in the $\phi N$ -
subsystem ($E_{\phi N}\sim-9$ MeV) one can expect that in the 
3-particle $\phi n p$ system the configuration $\phi+d$ is rather
suppressed. Hence, it follows that in the above system there is
no strong cancellation between potential and kinetic energies of
nucleons, like in deuteron, and a strong attractive triplet $N-N$
potential ($V_{\,t}\sim 100$ MeV) shows its full value.

\begin{figure}%
\begin{center}
{\includegraphics[angle=0,width=11cm]{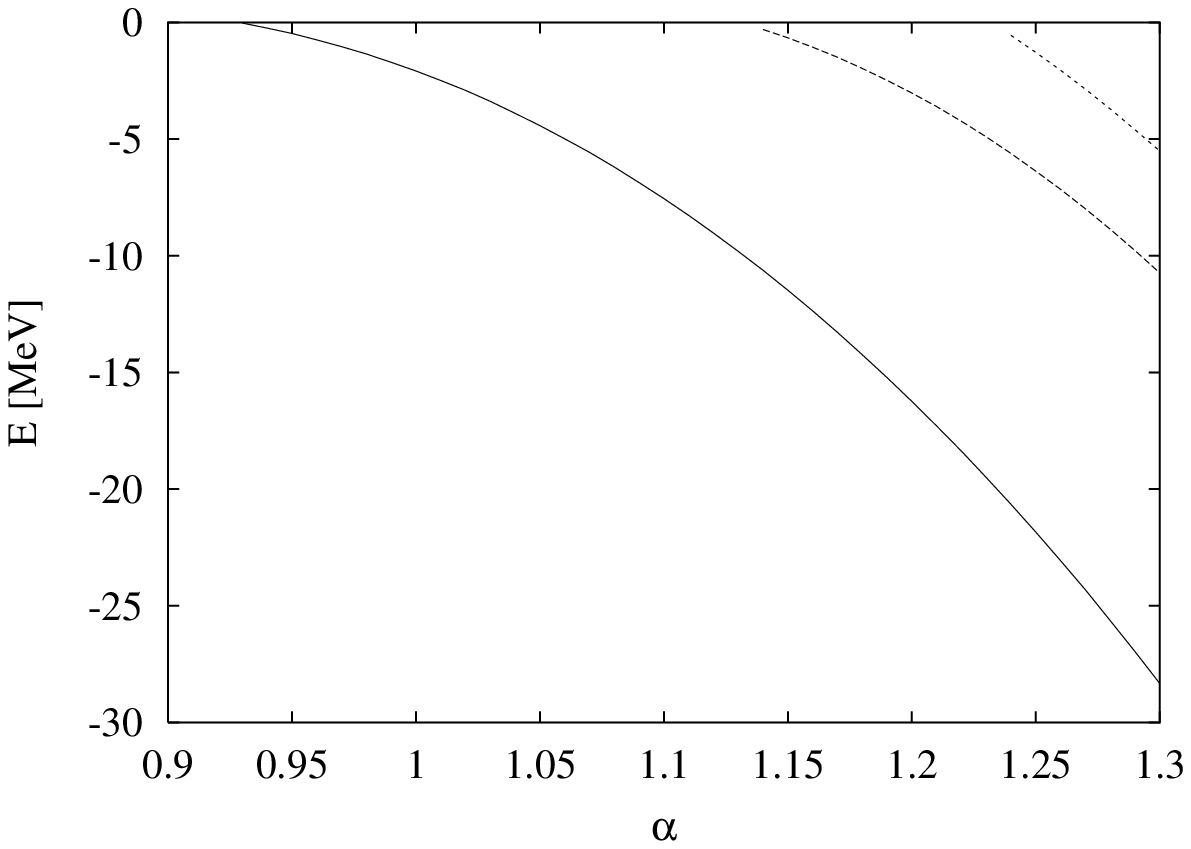}}%
\quad\mbox{}
\end{center}
\vspace{-7mm} \caption{The dependence of the binding energy of the
$\phi n n$ system on the parameter $\alpha$ of the $\phi-N$
interaction.} \label{phi+n+n ot alpha}
\end{figure}

The dependence of the $\phi n n$ binding energy on the parameter
$\alpha$ of the $\phi-N$ interaction is investigated. It is shown
in Figure \ref{phi+n+n ot alpha} that excited states appear in
this system.

As can be seen from these results, the binding in 3-particle systems
like $\phi N N$ is possible even for weaker $\phi-N$ attraction
than the one of
 potential~(\ref{Yukawa phi N}).

Now let us turn to the $\phi \phi N$ system. The indication of a
resonance in the $\phi + \phi$ subsystem and its parameters have
been found by drawing the Argan diagram and studying the position
of the pole in the scattering amplitude.

 With the potential $V_{\phi \phi}$ of the form~(\ref{phi-phi potential}), the binding energy
 of the
 3-body system $\phi \phi n$ turns out to be - 77 MeV.

As in the case of the $\phi(np)_{triplet}$ state, due to the strong
 binding in the $\phi \phi N$ system, the K-meson decay channels of the $\phi$ - meson
 are closed.

 Let us discuss now the 4-body system $\phi$nnn.
 To make a preliminary estimate of its binding energy, we
 have used the folding model with the ($\phi n n$)-subsystem as a
 cluster.

\begin{figure}%
\begin{center}
\includegraphics[angle=0,width=11cm]{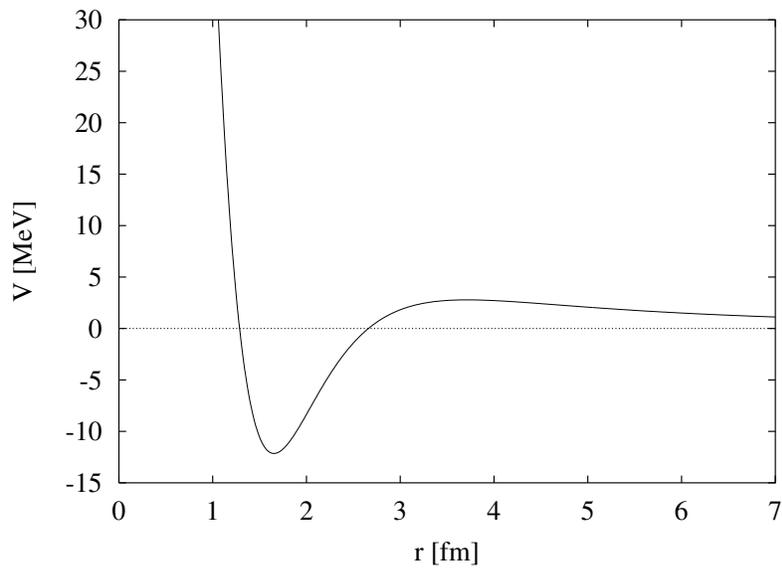}
\end{center}
\vspace{-7mm} \caption{The folding potential with the p-wave centrifugal
barrier for the four-body system~($\phi n n)\,+\,n$.}
\label{folding plot}
\end{figure}

 By averaging the interactions of the third neutron with the particles of the ($\phi n n$)
 system over the cluster
 wavefunction
 it is easy to obtain an effective potential which has the
 form shown in Figure \ref{folding plot}. Here $r$ means the distance between the
 third
 neutron and the center of mass of the ($\phi n n$) cluster.

  The solution of the Schroedinger equation with this potential shows
  that there are no bound states. However, the
  exact treatment of an analogous 4- body system ($\eta_c + 3N$),
  performed by means
 of the AGS- equations~\cite{Belyaev Shevchenko Fix Sandhas},
  shows that the folding model may
  strongly underestimate the exact calculation.

   We plan to perform such a calculation
  in the near future.

\begin{figure}%
\begin{center}
\includegraphics[angle=0,width=9cm]{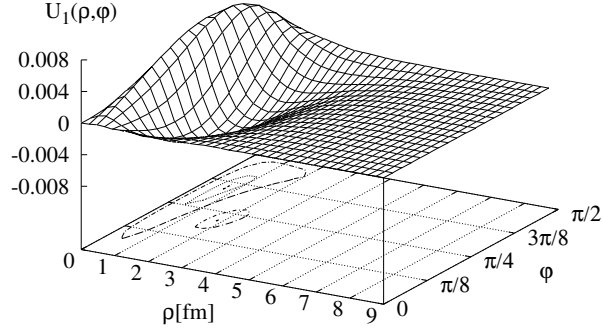}
\end{center}
\vspace{-7mm} \caption{The partial amplitude $U_1(\rho, \varphi)$
for the~$\phi n n$ system. Extreme values of $U_1(\rho, \varphi)$ are $6.772\times10^{\,-3}$ (at $\rho=0.654$, $\varphi=1.092$) and $-5.633\times 10^{\,-3}$ (at $\rho=1.881$, $\varphi=0.695$). Contours of different cross-sections of the $U_1(\rho, \varphi)$ surface are shown in the $(\rho, \varphi)$ plain. Dotted, dashed-dotted and second dashed-dotted lines correspond to values $5.643\times10^{\,-3}$, $1.693\times 10^{\,-3}$,  $-5.121\times 10^{\,-3}$ on the $U_1(\rho, \varphi)$ axis respectively.} \label{U1}
\end{figure}

\begin{figure}%
\begin{center}
\includegraphics[angle=0,width=9cm]{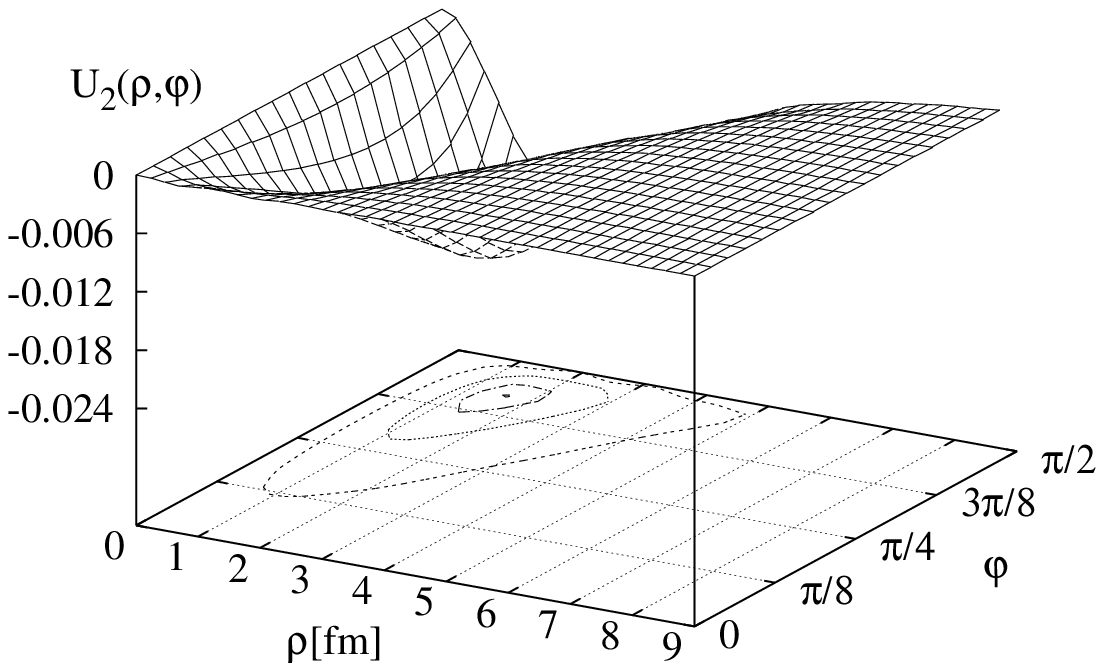}
\end{center}
\vspace{-7mm} \caption{The partial amplitude $U_2(\rho, \varphi)$
for the~$\phi n n$ system. The extreme value of $U_2(\rho, \varphi)$ is $-2.182\times10^{\,-2}$ (at $\rho=1.554$, $\varphi=1.323$). Contours of different cross-sections of the $U_2(\rho, \varphi)$ surface are shown in the $(\rho, \varphi)$ plain. Dashed-dotted, dotted and dashed lines correspond to values $-1.983\times10^{\,-2}$, $-1.454\times10^{\,-2}$ and $-0.545\times10^{\,-2}$ on the $U_2(\rho, \varphi)$ axis respectively.} \label{U2}
\end{figure}

Finally, let us discuss the matter distribution, shown in Figures
\ref{U1} and \ref{U2}, for the $\phi n n$ system. First of all, as
one can see on both the pictures which show the profiles of two
Faddeev components of the wave function, the main part of a matter
is concentrated at small distances inside the volume of the size
of 2 fm. In addition, one can see that the fine
structure of different components of the wave function is nonhomogeneous. Indeed,
the component $U_{\,2}$ has an extremum at hyperangle
$\varphi\;\sim\;\pi / 2$ which means that a $\phi$- meson and one of neutrons are close to
each other forming the substructure in the $\phi n n$ system, which 
was mentioned at the end of the section 2. In contrast to that, the component of the
wave function $U_{\,1}$ has two extrema without well- pronounced substructures.
 Having in mind the small size of the system and its neutrality,
 one would expect the propagation of an object like this far away from
 the region of production.
 
\section*{Acknowledgments} The work was supported by the
Deutsche Forschungsgemeinschaft (DFG grant no 436 RUS 113/761/0-2)
and the Heisenberg-Landau Program 2008.

\newpage

\begin{thebibliography}{99}
\bibitem{Weise} Dot\'e, A., Hyodo, T., Weise, W.: $K^-pp$ system with chiral SU(3) effective interaction. ArXiv:
nucl-th/0802.0238 (2008)
\bibitem{Yamazaki} Yamazaki, T., Akaishi, Y.: ($K^-,\pi^-$) production of nuclear $\overline{K}$ bound states in proton-rich systems via $\Lambda^*$ doorways. Phys. Lett. B 535,
70 (2002)
\bibitem{Agnello} Agnello, M., et al.: Evidence for a Kaon-Bound State $K^-pp$ Produced in $K^-$ Absorption Reactions at Rest. Phys. Rev. Lett. 94, 212303 (2005)
\bibitem{Oset} Magas, V. K., Oset, E., Ramos, A., Toki, H.: Critical view on the deeply bound $K^-pp$ system. Phys.
Rev. C74, 025206 (2006)
\bibitem{Yamazaki2} Yamazaki, T., Akaishi, Y.: Basic $\overline{K}$ nuclear cluster, $K^-pp$, and its enhanced formation in the $p+p \rightarrow K^+ + X$ reaction. Phys. Rev. C76, 045201 (2007)
\bibitem{Nina Gal} Shevchenko, N. V., Gal, A., Mare\v s, J.: Faddeev Calculation of a $K^-pp$ Quasibound State. Phys.
Rev. Lett. 98, 082301 (2007)
\bibitem{Nina Revai} Shevchenko, N. V., Gal, A., Mare\v s, J., R\'evai,
J.: $\overline{K} N N$ quasibound state and the $\overline{K} N$ interaction: Coupled-channels Faddeev calculations of the $\overline{K} N N - \pi \Sigma N$ system. Phys. Rev. C76, 044004 (2007)
\bibitem{Huang} Huang, F., Zhang, Z. Y., Yu, Y. W.: Baryon-meson interactions in chiral quark model. ArXiv:
nucl-th/0601003 (2006)
\bibitem{Gao} Gao, H., Lee, T.-S. H., Marinov, V.: $\varphi - N$ bound state. Phys. Rev. C63,
022201 (2001)
\bibitem{Tjon} Malfliet, R. A., Tjon, J. A.: Solution of the Faddeev equations for the triton problem using local two-particle interactions. Nucl. Phys. A127,
161-168 (1969)
\bibitem{Howell} Howell, C. R., et al.: Toward a resolution of the neutron-neutron scattering-length issue. Phys. Lett. B444, 252-259 (1998)
\bibitem{our in ArXiv} Belyaev, V. B., Sandhas, W., Shlyk, I. I.: New nuclear three-body clusters $\varphi NN$.
ArXiv: nucl-th/0707.4615 (2007)
\bibitem{Faddeev} Faddeev, L. D.: Scattering theory for a three particle system.
Zh. Eksp. Teor. Fiz., 39, 1459 (1960) (Sov. Phys. JETP, 12, 1014,
1961)
\bibitem{Pupyshev} Pupyshev, V. V.: On three-particle integrodifferential equations.
Theor. Math. Phys., 81:1, 1072-1077 (1989)
\bibitem{Belyaev Shevchenko Fix Sandhas} Belyaev, V. B., Shevchenko, N. V., Fix, A. I., Sandhas,
W.: Binding of charmonium with two- and three-body nuclei. Nucl. Phys. A780, 100-111 (2006)
\end{thebibliography}
\end{document}